# First Breakthrough for Future Air-Breathing Magneto-Plasma Propulsion Systems

**B Göksel[1*], I Ch Mashek[2]**

[1] Electrofluidsystems Ingenieurbüro Göksel, Berlin, Germany
[2] St. Petersburg State University, St. Petersburg, Russia

[*]Email: berkant.goeksel@electrofluidsystems.com

**Abstract.** A new breakthrough in jet propulsion technology since the invention of the jet engine is achieved. The first critical tests for future air-breathing magneto-plasma propulsion systems have been successfully completed. In this regard, it is also the first time that a pinching dense plasma focus discharge could be ignited at one atmosphere and driven in pulse mode using very fast, nanosecond electrostatic excitations to induce self-organized plasma channels for ignition of the propulsive main discharge. Depending on the capacitor voltage (200-600 V) the energy input at one atmosphere varies from 52-320 J/pulse corresponding to impulse bits from 1.2-8.0 mNs. Such a new pulsed plasma propulsion system driven with one thousand pulses per second would already have thrust-to-area ratios (50-150 kN/m²) of modern jet engines. An array of thrusters could enable future aircrafts and airships to start from ground and reach altitudes up to 50km and beyond. The needed high power could be provided by future compact plasma fusion reactors already in development by aerospace companies. The magneto-plasma compressor itself was originally developed by Russian scientists as plasma fusion device and was later miniaturized for supersonic flow control applications. So the first breakthrough is based on a spin-off plasma fusion technology.

**Introduction**

Magneto-plasma compressors (MPC) were originally developed as quasistationary plasma accelerators for fusion applications [1, 4]. They consist of coaxial electrodes representing an electromagnetic Laval nozzle analogue [4, 8], see Figure 1. In the air regime with initial pressures from 0.1-10 Torr the plasma jet has a lifetime of 50-100 μs, can reach velocities from 5-20 km/s and high densities from $10^{17}$-$10^{18}$ cm$^{-3}$ [8]. In the plasma focus the pinching discharge can be compressed to pressures of about 100-150 bar [8]. High speed plasma jets generated by magneto-plasma compressors (MPC) can be applied for plasma fusion devices [1-6, 15, 26-28], high-enthalpy plasma-material interactions, surface modification [7, 15, 25, 27], supersonic flow control [8-11] and air-breathing propulsion [12-14].

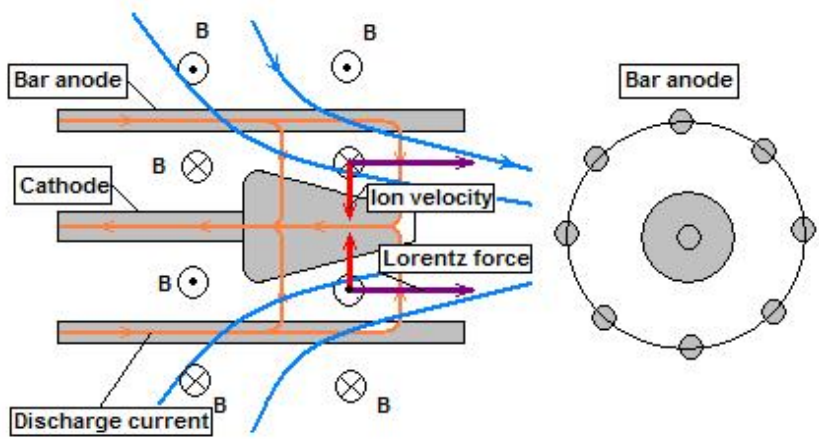

**Figure 1.** Physical principle of magneto-plasma compressors as electromagnetic Laval nozzle analogue [8].

**Motivation**

Almost all known MPC types operate under high vacuum conditions ($10^{-5}$-10 mbar) with very high discharge currents (10-200 kA) and low repetition frequencies (0,01-0,1 Hz). The overall system in all cases is based on bulky high-current switchers, heavy high voltage capacitor banks with tens of kV, complex control and diagnostics devices [1-7]. Nevertheless, all the previous work gave the main inspiration to develop new lightweight, highly miniaturized, lower voltage powered stable MPC-based plasma jet generator systems for flow control and propulsion purposes. The main initiation was given in 2010 by an informal request from an European aerospace company for a stratospheric high-thrust plasma propulsion with tens to hundreds of Newton. At the time this technology was not available and the request could not be addressed but the idea was born to develop a new propulsion system which combines electrohydrodynamic barrier discharge and field electron emission effects with magnetohydrodynamic flux compression aspects. The first breakthrough milestone goal was to realize a MPC system with 5-10 Hz repetition frequency which is able to operate under atmospheric pressures from 0.1-1.0 bar [13]. So far there was no known work about MPC operating at high-pressure and high-frequency. This challenge and the industrial request for a high-altitude propulsion systems was the main motivation to start a R&D programme for pulsed plasma "detonation" thrusters.

It is not widely known that a hydromagnetic Rankine-Hugeniot model for detonation and deflagration can be also used to describe gas-fed (air-breathing) coaxial plasma accelerators which are based on pinching dense plasma focus discharge devices like the magneto-plasma compressor. The gas is in this case is resistively heated by the pinching discharge instead of the burning in the combustor. The hydromagnetic shock caused by the high magnetic pressure ionizes and compresses the gas volume ahead of itself [19-22]. Nevertheless, all the known studies focus on high-vacuum conditions with plasma jet velocities exceeding 100-200 km/s.

Recent studies about sub-millimeter dense plasma focus (DPF) devices also reveal attempts and proposals towards higher pressures of 10-1000 Torr in hydrogen though limitations in confinement due to increased collisions at high pressures are expected [23]. Nevertheless, the same authors could recently demonstrate a DPF operating in 50-190 Torr Helium [24]. This specific type of pinching plasma accelerator is producing short nanosecond discharge pulses which are also relevant for x-ray and short pulse neutron emissions. DPF devices and their applications are again reconsidered for investigation in large-scale plasma fusion and space propulsion programs [26-31]. This is a further motivation to study air-breathing magneto-plasma jet thrusters.

In previous experiments the authors used microsecond discharges for internal initiation of the propulsive main discharge and could already increase the MPC operation pressure from 30 to 250 Torr (0.33 bar), see Figure 2a-2c [13]. The use of nanosecond discharge was the next logical step which was also motivated by the fact that in experiments to chemical pulse detonation engines distributed or transient nanosecond spark discharges are significantly more efficient for detonation initiation than localized microsecond spark discharges of comparable pulse energy [17]. So high repetition nanosecond pulsed discharge are already in the focus of groups working on efficient combustion processes for several years [18], though no working group has obviously ever combined the nanosecond know-how with pinching dense plasma focus discharges.

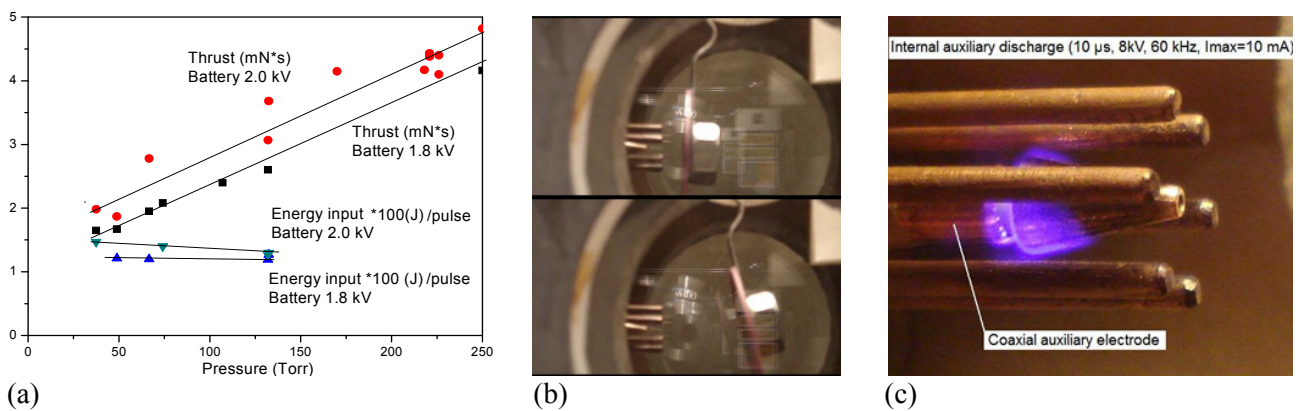

**Figure 2a-2c.** Thrust and pulse energy deposition in dependence of the air pressure for different battery voltages with internal microsecond sliding discharge exitation [13].

**General experimental setup**

The new MPC has an outlet diameter of 14 mm and contains six coaxial bar anodes (each from 3mm Copper rods) and one conical Copper cathode with a maximum diameter of 7 mm. The minimum distance between each anode and cathode is 2 mm. Following the electrode size and distances we can name this configuration (3-2-7) mm. The overall length of the new compressor is 80 mm, see Figure 3a. The MPC location inside an acrylic vacuum chamber is shown in Figure 3b and 3c.

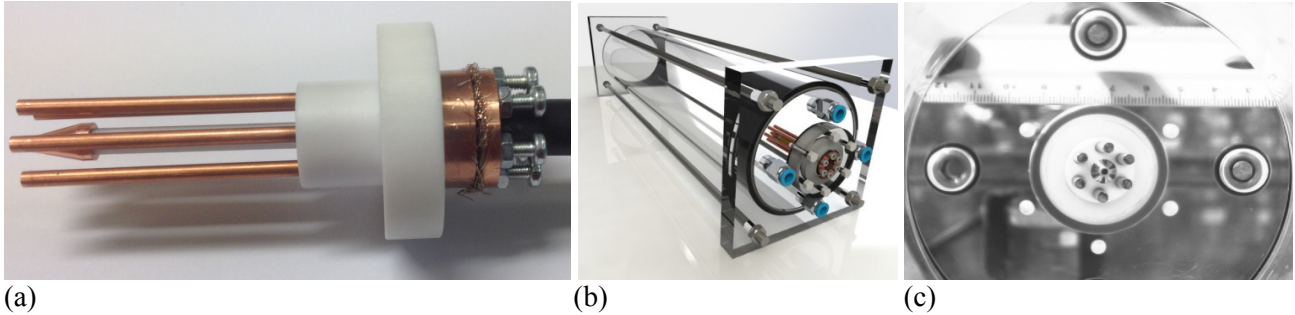

(a)　　　　　　　　　　　　　　　　(b)　　　　　　　　　　　　　　　　(c)

**Figure 3a-3c.** Tested MPC and test vacuum chamber.

The MPC is designed to work under pressures from 0.1-1.0 bar. The related self-breakdown threshold for the new (3-2-7) configuration is below 5kV. A nanosecond (ns) high voltage pulse generator (NPG-18/3500N, Megaimpulse Ltd, Russia) is used to induce a homogeneous, transient discharge for internal MPC excitation. High-frequency nanosecond pulse discharges have the unique property to induce transient self-organization of plasma channels between the electrodes [16, 17], see Figure 4a-4c. The transient ns-discharge is driven by -15 kV, 3.5 kHz pulses with 4 ns rise time and 10-20 ns pulse length for all test pressures. The full dynamic behavior of the ns-discharge will be addressed in a separate paper. The general scheme of experimental setup for the new MPC investigation is presented below in Figure 5.

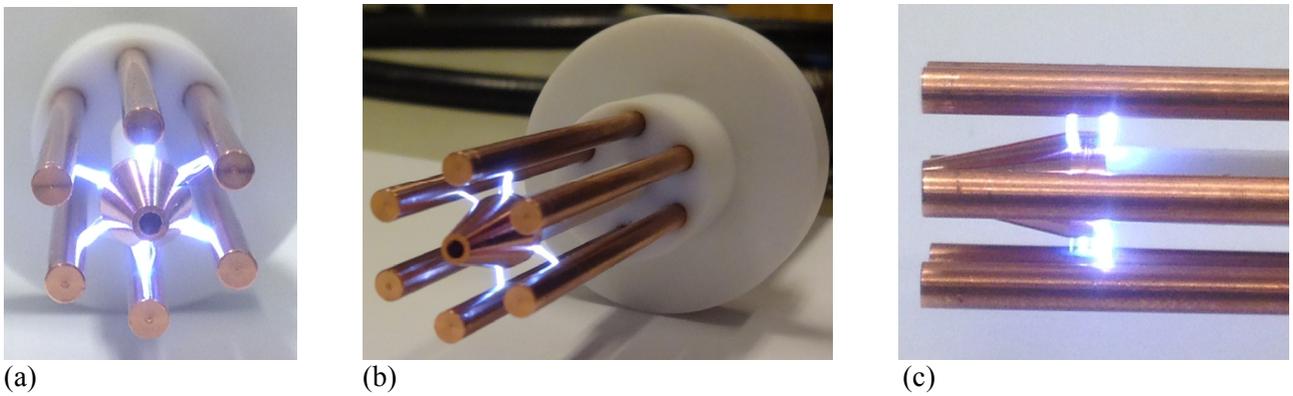

(a)　　　　　　　　　　　　　　　　(b)　　　　　　　　　　　　　　　　(c)

**Figure 4a-4c.** Photos of transient atmospheric nanosecond pulse discharges for internal MPC excitation.

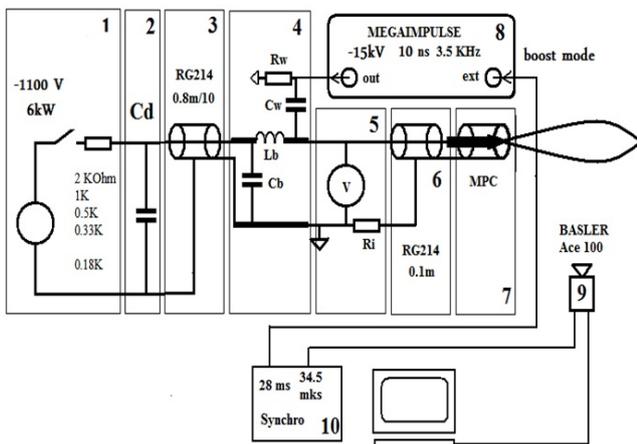

**Figure 5.** General scheme of the setup.

The new MPC (7) is connected to a cable collector (4) through a short piece of coaxial cable (6) (0.1 m RG214U) and a discharge current/voltage measuring circuit (5) (current resistor 0.75 mOhm, voltage sensor PMK-14KVAC). The cable collector contains a high-frequency blocking filter ($L_b$=150 nH, $C_b$=1,1 nF), a separating capacitor $C_w$ (220 pF, 16 kV), a pull-down resistor $R_w$ (75 Ohm) and a low-inductance Copper plate which leads the current for the capacitor bank (battery) output cables. The cable collector (4) itself is connected with the main discharge battery by ten coaxial cables (3) having each 0.8 m length (RG214U). The main discharge capacitor bank consists of 10 impulse capacitors with 2700 µF and 1200V maximum voltage. This capacitor bank (2) is charged through the resistor $R_c$ and a voltage regulator with a maximum output voltage of 1100 V and 6 kW power.

The nanosecond pulse generator (8) (NPG-18/3500N) operates in the boost regime (external trigger). The external synchronization input (28 ms) generates about 98 high voltage (HV) ns-pulses which again induce the homogeneous, transient plasma channels between the cathode and coaxial anodes. The blocking filter in the cable collector (4) saves the main battery from damage through each of the nanosecond HV pulses.

The capacitor bank (2) has a total capacity of 2700 µF, an internal resistance of 0.3 mOhm and an internal inductance of 8.5 nH (10 capacitors in parallel connection). The cables connecting the battery with the cable collector have a self-inductance of 20 nH and a resistance of 0.6 mOhm (10 RG214U cable with each 0.8 m in parallel). The cable collector has the resistance of 0.02 mOhm and a self-inductance of 6.6 nH. The inductance of the blocking filter is 150 nH. The short connecting cable for the MPC (0.1 m RG212U) has 25 nH and 0.7 mOhm.

**Visualization setup and results**
Visualization of MPC induced plasma jets was performed with a Basler Ace100 camera (9 in Figure 5) having a frame exposition able to synchronize with the nanosecond HV pulse generator working in a boost regime with 28 ms. The camera itself was launched with a synchronizing circuit (10 in Figure 5) using short TTL pulses with a duration of 34.5 µs. The end of a pulse launches the boost regime of the ns pulse generator (8 in Figure 5). To minimize EM interferences from the nanosecond pulse discharges to the camera and computer electronics the synchronization pulse was transferred to the nanosecond HV pulse generator through an optical fiber coupling.

In Figure 6a-6c, a general view of the high pressure plasma jet is presented. All pictures were taken with a dense red glass filter KS-15 and minimal camera lens diaphragm.

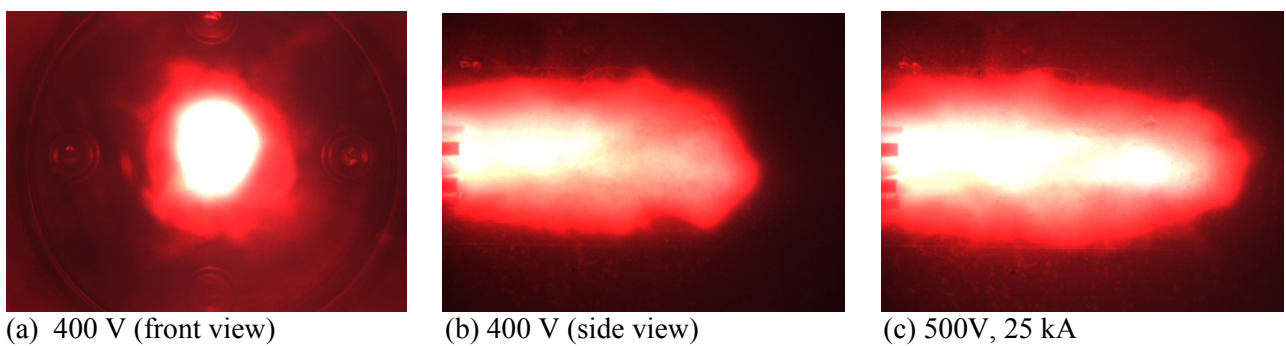

(a)  400 V (front view)    (b) 400 V (side view)    (c) 500V, 25 kA

**Figure 6a-6c.** View of MPC plasma jets for different battery voltages $U_b$ at an air pressure of 450 mbar.

Usually, the main discharge is arising after 2-10 pulses of the internal ns-excitation. At one atmosphere the MPC plasma jets have a similar character but higher intensity than under a lower pressure of 450 mbar, see Figure 7a-7c. In Figure 7a, it can be clearly seen that a pinching plasma jet with a compression focus area already exists for a battery voltage as low as 300 V. The pinching plasma can be even more clearly seen through IR and UV filters, see Figure 8a-8b. Figure 9a-9b show a visualization of the nanosecond excitation (Figure 4a) and main discharge through a blue filter. In all previous studies the battery voltage was much higher in the range of at least 1.8 kV [13]. Nevertheless, the capacitor bank can provide voltages up to 1.2 kV.

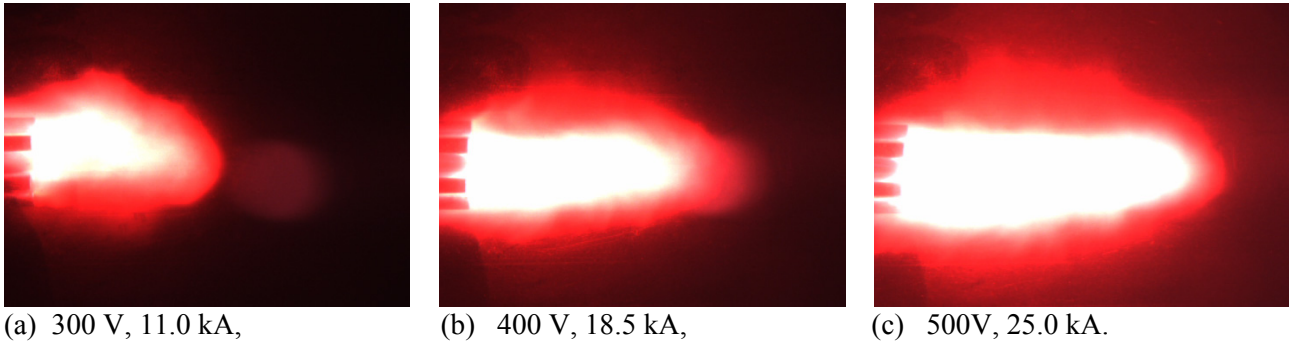

(a) 300 V, 11.0 kA,   (b) 400 V, 18.5 kA,   (c) 500V, 25.0 kA.

**Figure 7a-7c.** Plasma jets for different battery voltages $U_b$ at 1 bar atmospheric pressure through a red filter.

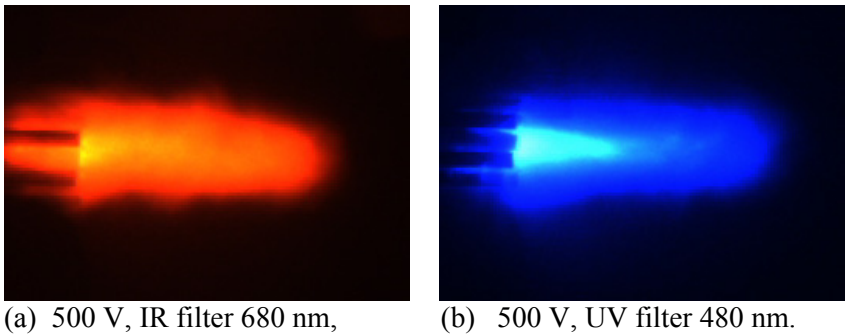

(a) 500 V, IR filter 680 nm,   (b) 500 V, UV filter 480 nm.

**Figure 8a-8b.** Plasma jet for $U_b$=500 V at 1 bar atmospheric pressure through IR and UV filters.

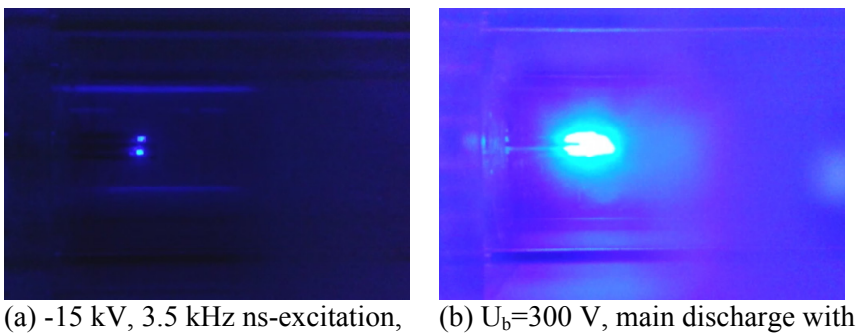

(a) -15 kV, 3.5 kHz ns-excitation,   (b) $U_b$=300 V, main discharge with $I_d$=11.0 kA peak current.

**Figure 9a-9b.** Excitation and ignition of main discharge at 1 bar atmospheric pressure through a blue filter.

**Current and voltage measurements and simulations**
The discharge current for the MPC has a weak dependence from pressure, and mainly depends from the discharge voltage. A typical presentation of the discharge parameters are shown in Figure 10a.

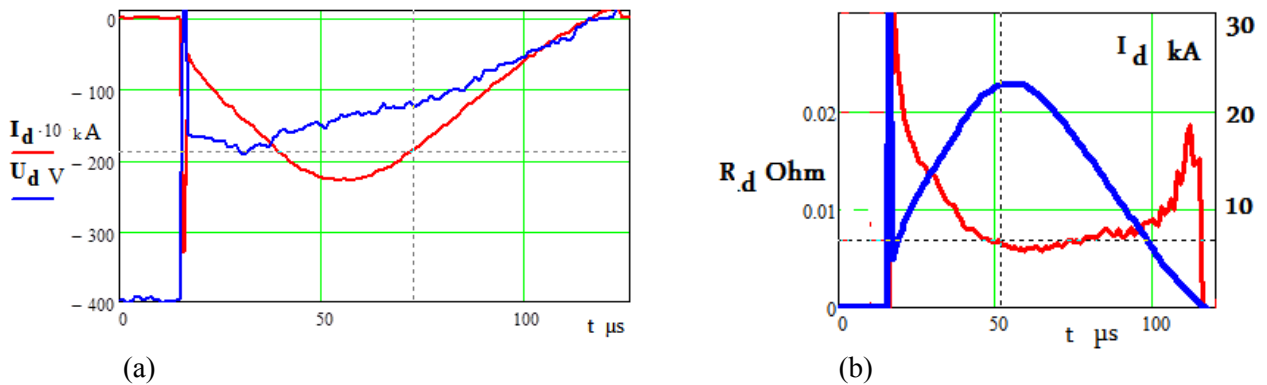

**Figure 10a and 10b.** MPC discharge for $U_d$=400 V (battery with 2700 µF) at 1 bar atmospheric pressure.

In Figure 10b, the nonlinear resistance (red line) of the pinching high-current arc discharge is shown over time. The discharge resistance $R_d$ dramatically decreases when the discharge current reaches the higher peak levels of 15-18kA in this specific case, Figure 10b. This effect is related to the tangential component of the strong magnetic field arising between MPC electrodes. The intrinsic tangential component of the magnetic flux density is the most important factor in this type of pinching plasma accelerators.

For a first estimation the value of the magnetic fields at different discharge currents were simulated in COMSOL, Figure 11a-11b. The electrodynamic model of the MPC contains six circuits with anode rods and a common center cathode. The plasma parameters were estimated from experimental data of current voltage characteristics. The average cross-section for each of the six plasma channels is about 4 mm² with a resistance of 60 mOhm. The current for each circuit is equal to 1/6 of the overall discharge current (25-30 kA).

The experimentally obtained relationship between the nonlinear discharge resistance and the discharge current clearly shows that the generation of a flux compression plasma jet in the MPC is only possible at a certain level of the tangential magnetic flux density in the discharge gap. Minimum 0.7-1.2 T is necessary for the investigated plasma accelerator. The distribution of the tangential component of the magnetic flux density in the electrode gap for a discharge current of 30 kA is shown in Figure 11a for a cross-section placed 1 mm downstream of the plasma channels. Figure 11b shows the radial distribution of the tangential component of the magnetic field for different discharge currents (direction cathode center to anode center).

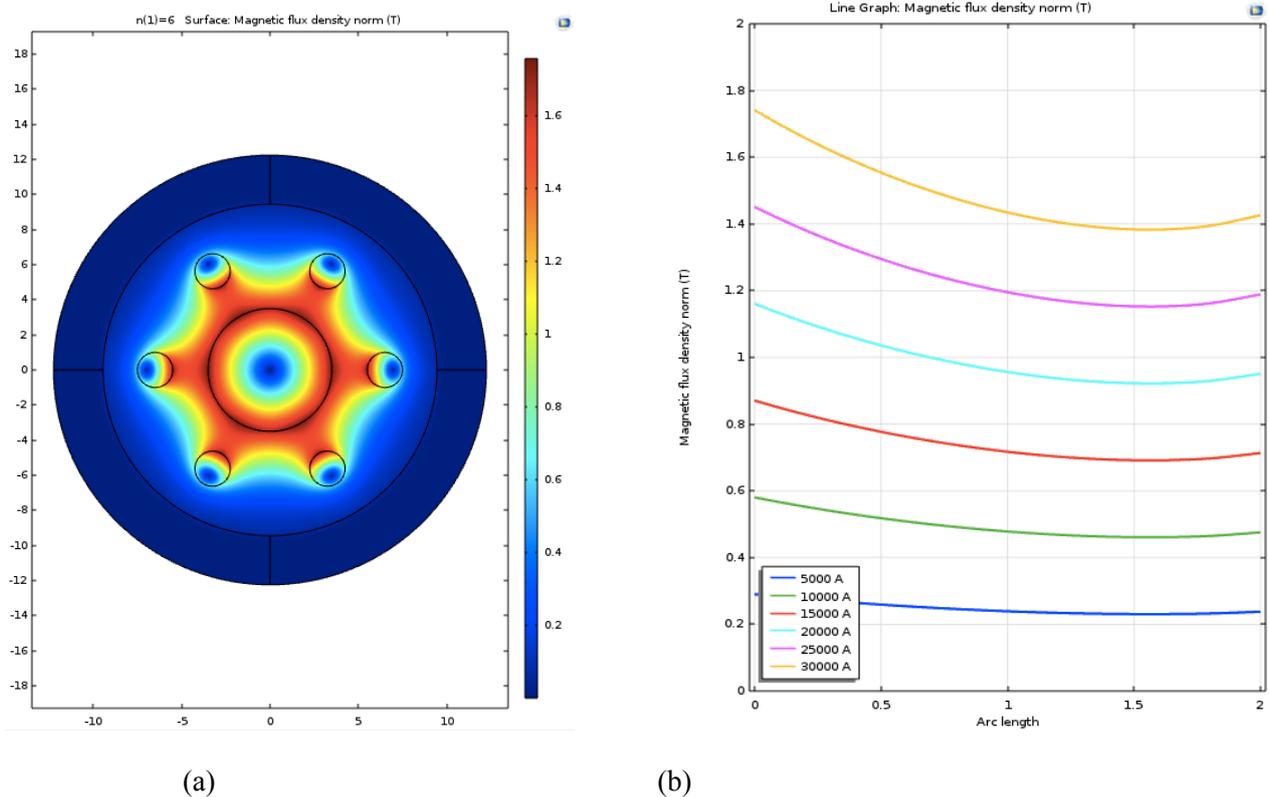

(a)          (b)

**Figure 11a and 11b.** The distribution of tangential magnetic fields between the MPC electrode gap.

The results of simulations and experiments show that there are strict requirements on the geometry of the discharge electrode. A flux compression with a low energy loss can be reached for discharge currents higher than 15-18 kA. This level corresponds to a minimum tangential magnetic field of 0.7-0.9 T. Particularly, the discharge voltage in the working gap at the proper geometry becomes so small that it causes low energy losses. Thus, the energy efficiency (ratio between discharge energy and energy of the capacitor bank) for the (3-2-7) mm MPC version with six-anodes at low voltage is about 0.85-0.90 but is decreasing for higher voltages.

The capacitor bank voltage $U_b$ here was varied from 200 to 600 V, though the maximum possible value is 1200 V. The overall energy efficiency for other voltages can be estimated or calculated using the formula (1), the $W_d$, $U_d$ values from Figure 14 and $C_b = 2700\mu F$.

$$\eta = W_d / W_b = \int I_d(t) \cdot U_d(t)\, dt / [C_b \cdot U_b \cdot U_b / 2] \quad (1)$$

**Impulse bit measurements**

The impulse bit of the new MPC was investigated using a free pendulum with 55 mm length and a mass of 15 g (see Figure 12a-12b). The pendulum deflection angle varied from 5° to 25°. Such small angles allow the use of the simplest model for a mathematical pendulum. The method used for thrust measurement is comparatively simple but has several sources for systematic errors. The first is the kind of plasma jet-surface interaction (elastic, non-elastic). An absolutely inelastic interaction of the plasma jet with the pendulum surface was assumed. The second connects with the uncertainty of contact point for the jet-surface interaction because the direction of the generated plasma jet shows some fluctuations. And the third main reason for errors is the stochastic nature of the impulse arc discharge at high pressures of 0.2-1.0 bar. The estimated measurement accuracy is 10-15 %. Figure 12c shows a series of small motion pictures taken from a video for a variant with a more lightweight pendulum. Conductive and insulating pendula made the same impulse bit. Figure 13 shows the measured impulse bit for different battery voltages and different atmospheric pressures. The first results show a comparatively weak dependence between impulse bit and atmospheric pressure in the measured range from 175-1000 mbar.

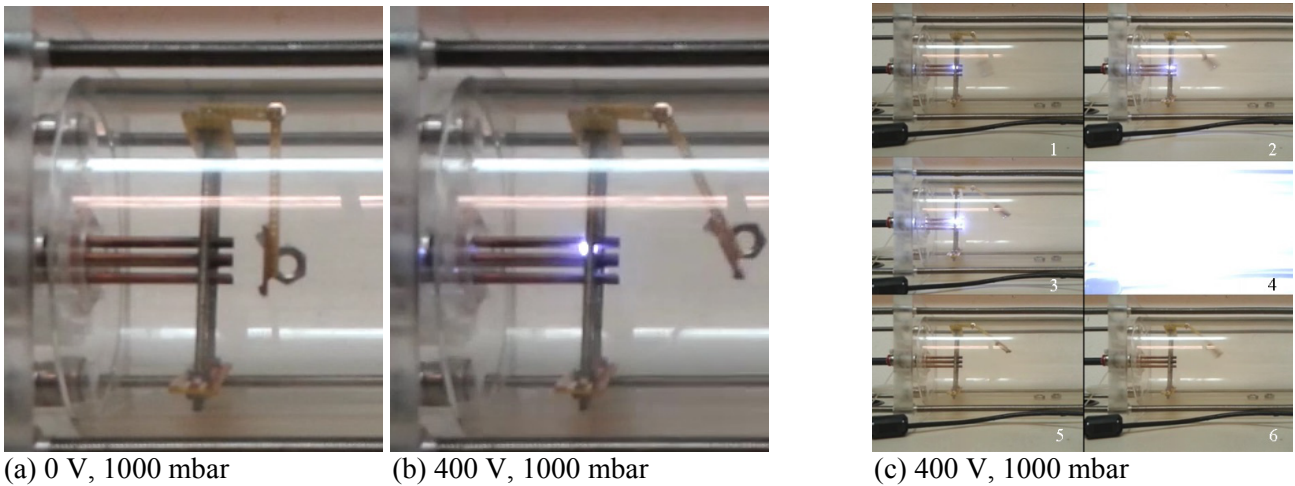

(a) 0 V, 1000 mbar     (b) 400 V, 1000 mbar     (c) 400 V, 1000 mbar

**Figure 12a-12c.** Measurement of impulse bit (Ibit) and thrust demonstration using a pendulum.

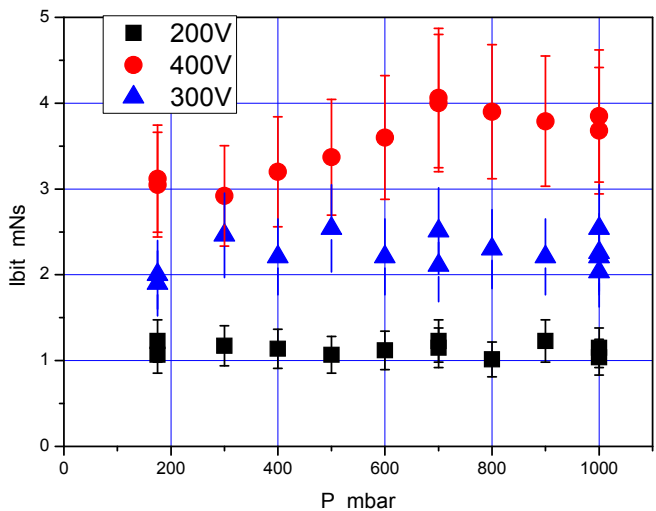

**Figure 13.** Measured impulse bit for different battery voltages and different atmospheric pressures.

**Summary and outlook**

Figure 14 gives a summary of the main results for the maximum discharge current $I_{d,max}$, the impulse bit and the discharge energy $W_d$. The input energy $W_b$ (proportional to the square of the battery voltage $U_b$) is dramatically increasing for higher input voltages, see also formula (1). The measured discharge energy $W_d$ is also increasing in a similar way. But at higher voltages beyond 400 V the energy loss is rising too. So the effective energy available to initiate the discharge is not proportional to the input energy. The kink in the discharge energy slope at 400 V is caused by the higher energy losses at 500 and 600 V. In these cases, the energy efficiency is only 0.74 and 0.66 compared to the much higher efficiencies at voltages below 500 V.

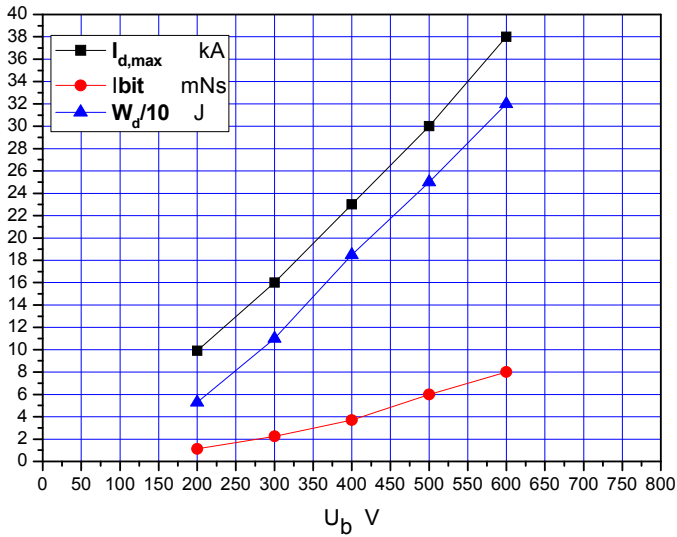

**Figure 14.** Summary of results for $I_{d,max}$, impulse bit and discharge energy $W_d$ for varying battery voltage $U_b$.

One of the most essential problems for future high-frequency MPC applications is the erosion of the electrode system. Figure 15 shows the results of the erosion process after approximately $10^3$ main discharge ignitions with a maximum current no more than 30 kA. The observed level of erosion was unexpectedly low as the studied high-current arc discharges with the self-induced strong tangential magnetic fields have fast moving cathode spots. But the main visual problem is in the diverter area (hole on the center of cathode) as shown in Figure 15. Possible ways for erosion decreasing in this area can be reached by changing the diverter diameter and shape. The theory and simulation of the plasma behavior in the MPC diverter area is very complicated. So future magneto-plasma simulations and experimental trials might give indications for a better diverter geometry with less erosion.

Nevertheless, future air-breathing magneto-plasma propulsion systems will need up to $10^3$ ignitions per second. So alternative materials from fusion reactors, amorphous metals and special alloys with different porosity and surface structuring will be be also investigated in the future. In this regards, the MPC itself might be useful to manipulate the surface structure of future electrode materials [7].

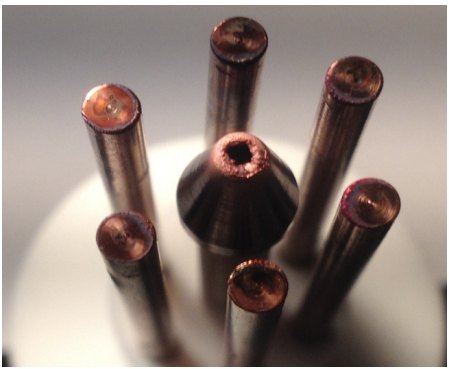

**Figure 15.** Results of erosion processes in MPC after $10^3$ launches.

In future experiments the plasma dynamics could be investigated using ultrafast cameras with up to 2 Mio frames per seconds which are available to the corresponding authors. In the present work the maximum voltage and power limits of the new MPC thruster were not tested. The main task was the first demonstration of a pulsed MPC-based plasma thruster with ns-internal excitation for a stabile operation at high atmospheric pressures up to 1 bar. In this regard, a first breakthrough and pulse operation with 4.7 Hz was demonstrated.

In the next step, the pulse frequency of the main discharge will be increased up to 10 Hz. Furthermore, a new mobile power generator will be developed for the first flight demonstration onboard of the b-Ionic Airfish, which was the world's first airship propelled by plasma engines in 2005, see Figure 16a-16b. Only 0.08 N or 8 g would be sufficient to propel this 7.5 m airship at low speeds up to 1 m/s [32]. A 5 Hz thruster has already about 0.02 N. So an array of four plasma pulse "detonation" thrusters with the present power level would make it fly. The available maximum weight for the power generator is about 5.1 kg plus 1.2 kg for LiPo batteries [32].

The general thrust of an array with 10 cells, each operating with a pulse frequency of 50 Hz, is 2.0 N. The total power required for a first high altitude (H=20 km) demonstrator mission using an array with 10 thruster is about 75 kW. With a solar battery effectiveness of 0.2-0.3, the required minimum surface area is 250 m².

In any case, the new propulsion technology is still away from being competitive but it has to be noted that the research, development and optimization is now at the very beginning. The impulse bit for each thruster unit can be essentially increased by using different ejector schemes and jet focusing nozzle structures. These are items of next investigations.

Furthermore, there are also a large amount of other possible technological applications in the field of aerodynamics, material sciences and power engineering. But a real flight demonstration is the next milestone goal towards new magneto-plasma flux compression thrusters for stratospheric airships or high altitude platform stations (HAPS) which are currently all limited to about 25 km altitude by using propellers. With future air-breathing magneto-plasma flux compression thrusters next generation solar, beamed or fusion energy powered airships could climb to altitudes up to 50 km and beyond.

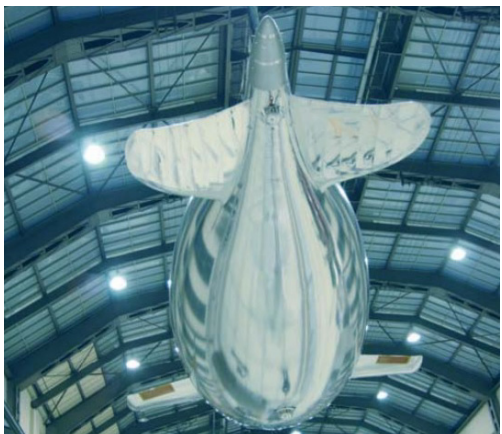
(a)
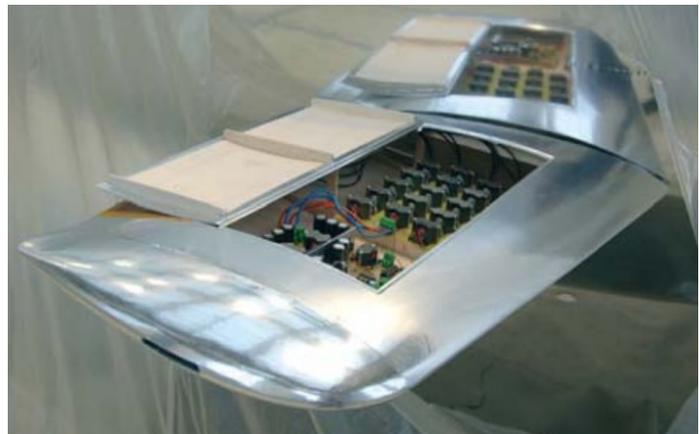
(b)

**Figure 16a and 16b.** The world's first plasma propelled airship, the 7.5 m long b-Ionic Airfish (2005) [32].

**Acknowledgements**
The breakthrough experiments were funded by Electrofluidsystems and performed by the authors in the company's future workshop in Berlin. The numerical studies were performed by Tatiana Banokina in St. Petersburg funded by a grant of the St. Petersburg State University with number 11.37.167.2014.